\documentclass[10pt,twocolumn,letterpaper]{article}

\usepackage{cvpr}              % To produce the CAMERA-READY version
\usepackage{amssymb}
\usepackage{fancyhdr}
\usepackage{times}
\usepackage{epsfig}
\usepackage{graphicx}
\usepackage{amsmath}
\usepackage{booktabs}
\usepackage{multirow}
\usepackage{multicol}
\usepackage{makecell}
\usepackage{color}
\usepackage{xcolor}
\usepackage{booktabs,tabularx}
\usepackage{bbding}
\usepackage{caption}
\usepackage{subcaption}
\usepackage{appendix}

\usepackage[pagebackref,breaklinks,colorlinks]{hyperref}
\usepackage[capitalize]{cleveref}
\crefname{section}{Sec.}{Secs.}
\Crefname{section}{Section}{Sections}
\Crefname{table}{Table}{Tables}
\crefname{table}{Tab.}{Tabs.}

\begin{document}

\title{Parameter-Efficient Transfer Learning for Audio-Visual-Language Tasks}

\author{
    %Authors
    Hongye Liu\textsuperscript{\rm 1,2},
    Xianhai Xie\textsuperscript{\rm 1},
    Yang Gao\textsuperscript{\rm 1},
    Size Li\textsuperscript{\rm 1},
    Zhou Yu\textsuperscript{\rm 3}\thanks{Corresponding Author}\\
    \textsuperscript{\rm 1}Kuaishou Technology, MMU \\
    \textsuperscript{\rm 2}School of Mechanical and Electrical Engineering, China JiLiang University \\
    \textsuperscript{\rm 3}School of Computer Science and Technology, Hangzhou Dianzi University\\
    \normalsize{\{liuhongye, xiexianhai, gaoyang08, lisize\}@kuaishou.com,} \\ 
    \normalsize{\{yuzhou\}@hdu.edu.cn}
}

\maketitle

\begin{abstract}
The pretrain-then-finetune paradigm has been widely used in various unimodal and multimodal tasks. However, finetuning all the parameters of a pre-trained model becomes prohibitive as the model size grows exponentially. To address this issue, the adapter mechanism that freezes the pre-trained model and only finetunes a few extra parameters is introduced and delivers promising results. Most studies on adapter architectures are dedicated to unimodal or bimodal tasks, while the adapter architectures for trimodal tasks have not been investigated yet. This paper introduces a novel Long Short-Term Trimodal Adapter (LSTTA) approach for video understanding tasks involving audio, visual, and language modalities. Based on the pre-trained from the three modalities, the designed adapter module is inserted between the sequential blocks to model the dense interactions across the three modalities. Specifically, LSTTA consists of two types of complementary adapter modules, namely the long-term semantic filtering module and the short-term semantic interaction module. The long-term semantic filtering aims to characterize the temporal importance of the video frames and the short-term semantic interaction module models local interactions within short periods. Compared to previous state-of-the-art trimodal learning methods pre-trained on a large-scale trimodal corpus, LSTTA is more flexible and can inherit any powerful unimodal or bimodal models. Experimental results on four typical trimodal learning tasks show the effectiveness of LSTTA over existing state-of-the-art methods. % Done
\end{abstract}

\begin{figure}[t]
    \begin{center}
    \includegraphics[width=1.0\linewidth] {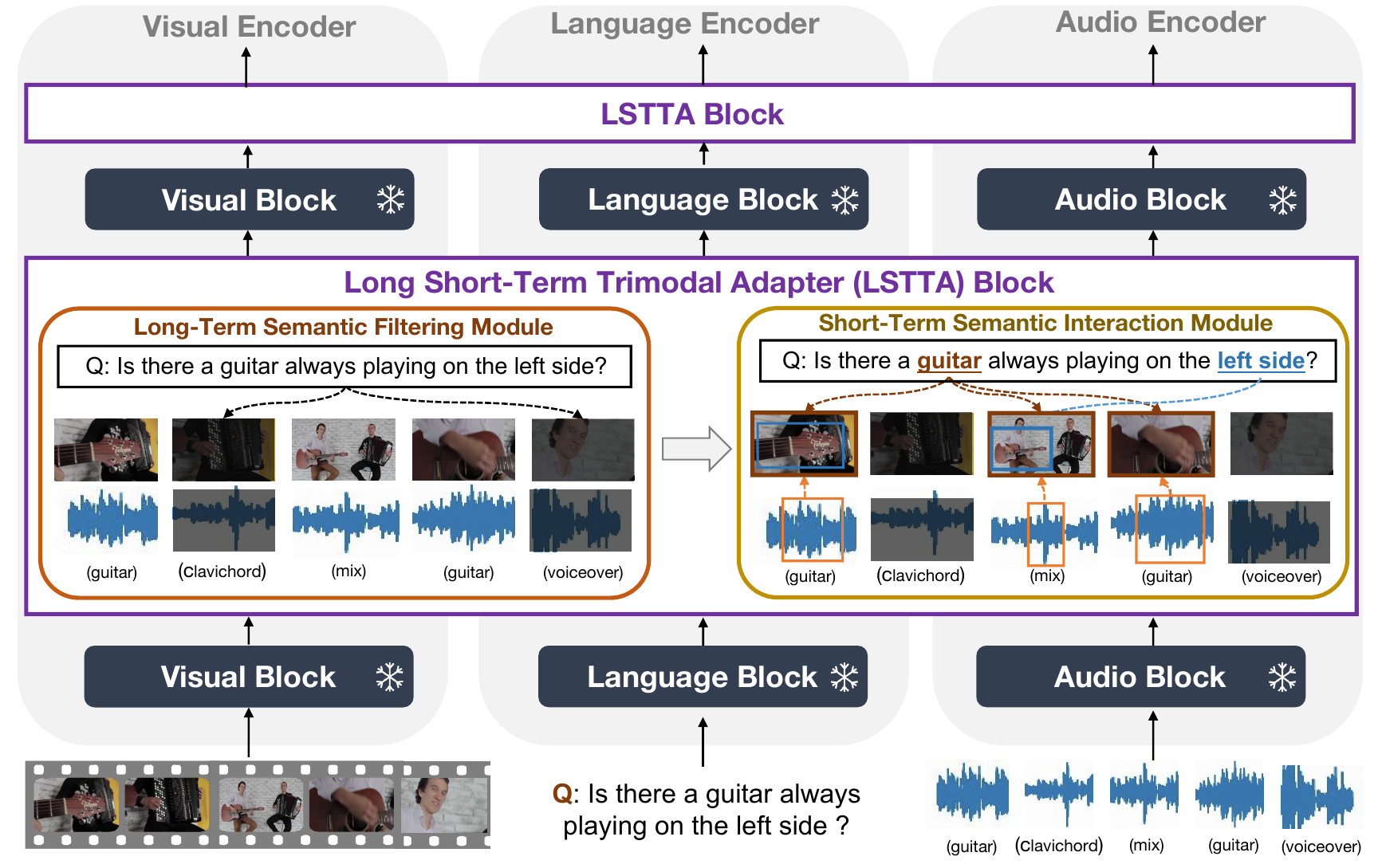}
    \end{center}
    \caption{
    The proposed LSTTA model consists of long short-term trimodal adapter blocks. Each block consists of two complementary modules. The long-term semantic filtering calculates the semantic importance over the whole video and suppresses redundant or less important frames. The short-term semantic interaction module characterizes the short-term semantic alignments across different modalities.% Done
    }
    \label{introduction}
\end{figure}

\section{Introduction}
In real-world scenarios, visual, language, and audio signals are commonly encountered and play a prominent role in many applications. Despite substantial advances in pre-training or transfer learning for one or two modalities, such as visual, visual-language, and audio-visual, effectively fusing and balancing all three modalities via transfer learning remains a significant challenge. The ability to interact and refine information from multiple modalities is crucial for the development of advanced artificial intelligence systems. %Done

Motivated by the success of the \emph{pretrain-then-finetune} paradigm in both the CV and NLP domains, there has been an increasing interest in transferring the paradigm to the multimodal domain and developing multimodal pretraining models to address a wide range of multimodal tasks. Recent studies on multimodal pretraining are focused on the visual-language domain. These visual-language pretraining (VLP) approaches first pretrain Transformer-based deep models on large-scale visual-language pairs and then finetune the pre-trained models on specific downstream tasks such as image-text retrieval~\cite{CLIP,FLAVA,ALBEF}, image captioning~\cite{BEiT3,VLMO,Uniter}, and visual question answering~\cite{BLIP,ALBEF}.  % Done

Despite that the VLP approaches have achieved promising results on various benchmarks, we argue that they have not taken full advantage of multimodal data. In addition to the visual and language modalities, we human beings also perceive information from the audio modality, which usually contains complementary semantics to visual and language modalities. However, adapting the prevailing VLP framework to support audio-visual-language tasks is non-trivial. %Done

In the realm of audio-visual-language (AVL) learning, prior studies have adopted the VLP paradigm to pretrain a trimodal model from scratch, heavily relies on a substantial amount of trimodal annotations derived from web videos~\cite{Audioclip,i-code,Vatt, OPT}. However, in reality, the availability of sufficient and diverse trimodal data is relatively limited compared to unimodal or bimodal ones, posing a significant obstacle to pretraining an AVL model adequately. This phenomenon raises a question: \emph{How to take full advantage of off-the-shelf pre-trained unimodal or bimodal model to adapt to the AVL scenario effectively and efficiently?}  % Done

One promising answer to this question is the parameter-efficient transfer learning mechanism ~\cite{Adapter,AdapterFusion}. This mechanism aims to adapt pre-trained models to new tasks by freezing the pre-trained model and inserting a few \emph{adapter} modules between every pair of layers in the pre-trained model. By introducing the adapter mechanism, the capacity of the pre-trained model is retained and the adaptation to new tasks is achieved efficiently. The adapter architectures are mainly dedicated to unimodal tasks and have achieved impressive results on various NLP~\cite{Adapter,AdapterFusion} and CV~\cite{VIT-Adapter,Conv-Adapter} tasks. Recently, several adapter approaches have been applied to the multimodal domain to address visual-language~\cite{Vl-adapter,Adapter-ju2022prompting} and audio-visual tasks~\cite{LAVISH}. % Done

To the best of our knowledge, the trimodal adapter architecture that incorporates audio-visual-language signals has not been investigated yet. In contrast to bimodal adapters, the design of trimodal adapters needs to address the following two challenges. First, the global semantics is not evenly distributed over time and some video snippets are more informative than others. Second, the audio and visual modalities have local correspondences within a short time period. Modeling the short-term semantic alignments without any annotation is crucial yet challenging. % Done

In this paper, we propose a long short-term trimodal adapter (LSTTA) method to address the above challenges of trimodal learning. The proposed LSTTA architecture consists of a sequence of LSTTA blocks, where each block consists of a long-term semantic filtering module and a short-term semantic interaction module. The long-term semantic filtering module leverages two non-parametric cross-modal attention blocks to aggregate long-term information from the visual and audio modalities, respectively. And then learns a temporal mask on top of them to suppress redundant video frames. LSTTA model consists of long short-term trimodal adapter blocks. The short-term semantic interaction module learns fine-grained semantic alignments across different modalities and facilitates local information transfer and aggregation across modalities. We evaluate the proposed method on four trimodal learning benchmarks: Music-AVQA, CMU-MOSEI, UR-FUNNY, and VIOLIN. Results show the superiority of LSTTA over existing state-of-the-art methods.% Done

The main contributions of this paper are threefold:
\begin{itemize}
    \item To the best of our knowledge, LSTTA is the first attempt to investigate the trimodal adapter architecture for audio-visual-language tasks. 

    \item LSTTA models long-term semantic importance and short-term semantic alignments simultaneously, which enhances the understanding of challenging AVL tasks. 

    \item Extensive experiments on three commonly-used public trimodal datasets show the advantage of our LSTTA method over existing state-of-the-art counterparts.
    
\end{itemize}

\section{Related Work}
In this section, we first provide a brief overview of visual-language models and audio-visual models. Next, we review the trimodal learning  tasks involving audio, visual, and language. Finally, we review the parameter-efficient transfer learning approaches, especially the adapter architectures for multimodal data. 

\subsection{Visual-Language \& Audio-Visual Models}
Most of current multimodal learning research focuses on visual-language tasks, e.g., visual question answering~\cite{FLAVA}, visual captioning~\cite{BEiT3,Flamingo}, and visual grounding~\cite{XVLM,X2VLM}. 
Early studies focus on a single task and design task-specific models~\cite{MCAN,DDPN,SCAN,lu2017knowing}. Motivated by the success of the pretrain-then-finetune paradigm of BERT in natural language understanding~\cite{BERT}, there has been an increasing interest in developing visual-and-language pretraining (VLP) models~\cite{Uniter,Oscar,VilBERT,Lxmert} to address a wide range of visual-language tasks. 
Early VLP approaches designed different pretraining tasks to learn multimodal Transformers on top of pre-extracted region-based visual features~\cite{Oscar,Ernie-vil,Rosita}. Motivated by the success of pre-trained visual backbones, e.g., ViT~\cite{ViT} and CLIP~\cite{CLIP}, recent VLP methods tend to exploit these visual backbones to obtain grid-based visual features and perform multimodal pretraining from the raw image and language inputs in an end-to-end manner~\cite{Ofa,ALBEF,BLIP,METER,shen2021much}. 

\begin{figure*}
    \begin{center}
    	\includegraphics[width=1.0\linewidth] {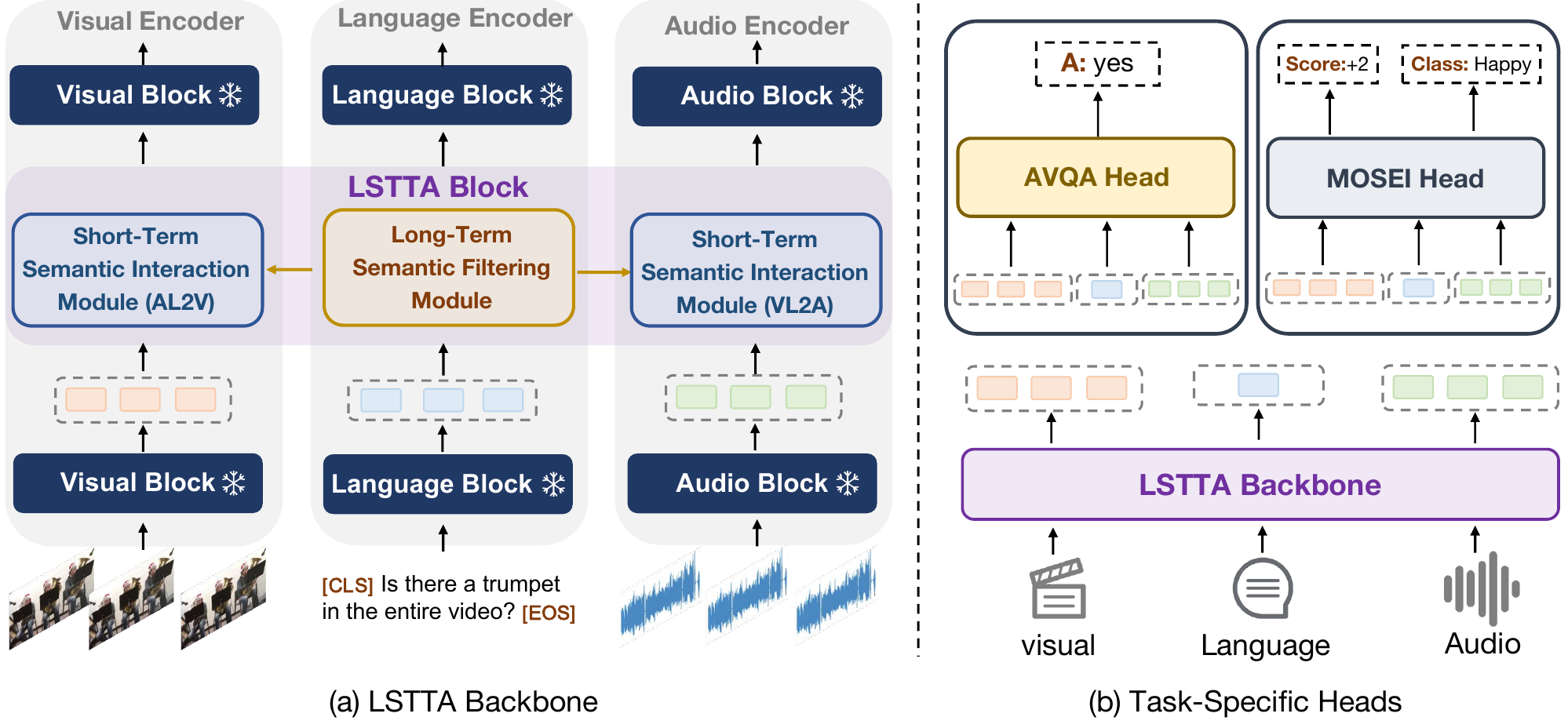}
    \end{center}
    \caption{The flowchart of the LSTTA framework, which consists of (a) the LSTTA backbone and (b) task-specific heads on top of the backbone for various trimodal learning tasks, namely Music-AVQA and CMU-MOSEI.}
    \label{network}
\end{figure*}

Audio-visual tasks aim at developing models that can effectively perceive and integrate both visual and audio features to comprehend activities in videos~\cite{Audio-Visual-Active,Audio-Visual-Contrastive,Audio-Visual-Cyclic,TVLT,LAVISH,AVE-CLIP,AVE-see,AVE-Dual}. Achieving this requires a robust mechanism for aligning visual and audio features, which is often accomplished by utilizing pre-trained models~\cite{AVE-tian-audio,AVE-lin2019dual,AVE-nagrani2021attention,CMBS} for feature extraction and designing fusion modules to combine the extracted features. %Done

\subsection{Audio-Visual-Language Trimodal Learning}
As the research on both visual-language and audio-visual tasks has made great progress so far, it is natural to investigate the audio-visual-language trimodal tasks. Recently, there has been increasing research on this direction~\cite{i-code,Merlot-reserve,Vatt,OPT}. For example, VATT employs a transformer encoder to project trimodal data from a unified backbone~\cite{Vatt}. i-Code utilizes merge-attention and co-attention mechanisms to integrate unimodal encoders~\cite{i-code}. AudioCLIP extends the VL model CLIP to AVL tasks by introducing an audio encoder based on ESResNeXt~\cite{Audioclip}. Despite the success of these methods, a common limitation of these methods is that they all need large-scale, parallel trimodal datasets to support sufficient training of their large models, which are often difficult to obtain in practice. As a consequence, how to maximally utilize pre-trained unimodal or bimodal models and devise an effective strategy to train models from limited trimodal data is a critical issue to be addressed.  %Done

\subsection{Parameter-Efficient Transfer Learning and Adapter Architectures}
Parameter-efficient transfer learning aims to adapt pre-trained models to new tasks by introducing a few trainable parameters~\cite{Adapter,Adapter-lin2022frozen,Adapter-liu2022polyhistor}.
The adapter mechanism is one of the most popular directions in parameter-efficient transfer learning, which introduces lightweight learnable modules inserted between every pair of layers in a pre-trained model. Most adapter approaches are dedicated to unimodal tasks, e.g., image classification~\cite{Tip-Adapter}, natural language understanding~\cite{Adapter,AdapterFusion}, and speech recognition~\cite{thomas2022efficient}. Recently, bimodal adapter approaches have been put forward. Lin \emph{et al.} introduce a LAVISH adapter method to handle audio-visual tasks, enabling the adaptation of frozen ViTs to these tasks~\cite{LAVISH}. Chen \emph{et al.} show that the standard transformer layer can serve as a good adapter architecture to extend pre-trained models from images to video~\cite{Adapter-ju2022prompting}. However, existing adapter architectures have primarily focused on one or two modalities, and balancing the fusion of information from three modalities in trimodal learning remains an open problem. To address this issue, we propose two complementary modules: a long-term semantic filtering, which characterizes the semantic importance of each frame over the entire video, and a short-term semantic interaction module, which models local interactions within a short time period.  

\section{The Proposed Method}
In this section, we describe the overall architecture of the proposed Long Short-Term Trimodal Adapter (LSTTA) method. We first overview the whole network architecture and then introduce the basic components of LSTTA. After that, we delve into the details of the two modules, namely the long-term semantic filtering and short-term semantic interaction modules. %Done

\begin{figure*}
    \begin{center}
    	\includegraphics[width=\linewidth] {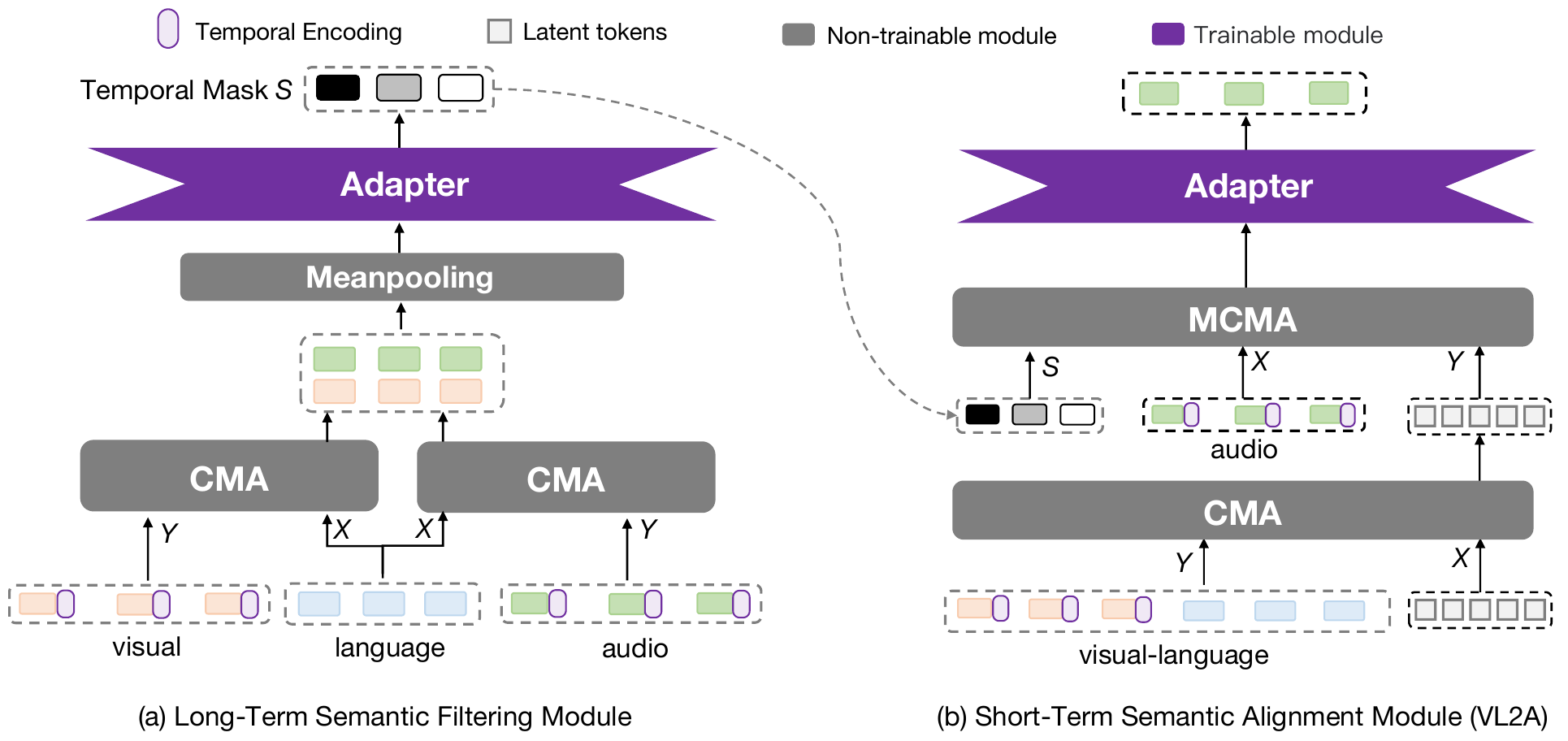}
    \end{center}
    % \vspace{-0.3cm}
    \caption{Illustrations of the two key modules in LSTTA, including (a) long-term semantic filtering (LTSF) and (b) short-term semantic interaction (STSI) module. For the STSI module, we show the VL2A version applied to the audio encoder. The AL2V version can be obtained by simply exchanging the places of the visual and audio features.}
    \label{fig:detail}
    %\vspace{-0.3cm}
\end{figure*}

\subsection{Overview}
\noindent\textbf{Network Architecture.} As shown in Fig. \ref{network}(a), the proposed LSTTA model contains multiple LSTTA blocks, where each block consists of one long-term semantic filtering (LTSF) module and two short-term semantic interaction (STSI) modules (i.e., AL2V and VL2A). Given pre-trained modality encoders from three modalities, each LSTTA block is inserted between two sequential encoder blocks across the three modalities.

Denote the visual encoder as $\mathcal{V}=\{V^{(1)},V^{(2)}..., V^{(L)}\}$, language encoder as $\mathcal{L}=\{L^{(1)},L^{(2)}..., L^{(L)}\}$, and audio encoder as $\mathcal{A}=\{A^{(1)},A^{(2)}..., A^{(L)}\}$, where each encoder consists of $L$ encoder blocks. Our LSTTA model is denoted as $\mathcal{M}=\{M^{(1)},M^{(2)}..., M^{(L-1)}\}$, which contains $L-1$ blocks\footnote{For the last encoder block, we do not append another LSTTA block after it.}. For each $j\in\{1,2,...L\}$, the procedures of the $j$-th LSTTA block can be expressed as follows:

\begin{equation}
\begin{split}
    z_v^{(j)}=V^{(j)}(f_v^{(j)}),~~
    z_a^{(j)}=A^{(j)}(f_a^{(j)}),~~
    z_l^{(j)}=L^{(j)}(f_l^{(j)}),\\
    [f_v^{(j+1)}, f_a^{(j+1)}, f_l^{(j+1)}] = M^{(j)}(z_v^{(j)},z_a^{(j)},z_l^{(j)}).
\end{split}\label{eq.1}
\end{equation}
where $f_v^{(j)}, f_a^{(j)}, f_l^{(j)}$ denote the input features of the $j$-th block of the visual, audio, and language encoders, respectively. $z_v^{(j)}, z_a^{(j)}, z_l^{(j)}$ correspond to the intermediate features of $f_v^{(j)}, f_a^{(j)}, f_l^{(j)}$, respectively. These intermediate features are further fed through the $j$-th LSTTA block $M^{(j)}(\cdot)$ to obtain the input features of the $j+1$-th encoder blocks.  %Done

As shown in Fig.~\ref{network}(b), on top of the last encoder blocks, the output features from the last encoder block of each modality are integrated and then fed to different task-specific heads to address various trimodal tasks, e.g., AVQA \cite{Music-AVQA} and MOSEI \cite{CMU-MOSEI}. 
\vspace{3pt}
\\
\noindent\textbf{Trimodal Representations.} Our trimodal learning tasks take video frames, text descriptions, and audio waveforms as inputs. For the visual encoder $\mathcal{V}$, it takes a sequence of $T$ video frames as its input and transforms each frame into $N_v$ tokens (patches). This results in the final visual features $f_v\in \mathbb{R}^{T \times N_v \times D_v }$, where $D_v$ is the feature dimensionality of each visual token. %Done

For the audio encoder $\mathcal{A}$, it takes $T$ audio waveforms as its inputs, where each waveform is temporally aligned to a certain video frame. After that, each waveform is encoded into $N_a$ audio tokens, resulting in the final audio features $f_a \in \mathbb{R}^{T \times N_a \times D_a}$, where $D_a$ is the dimensionality of each audio token. %Done

For the language encoder $\mathcal{L}$, it takes a sentence as its input and tokenizes the sentence into a maximum of $N_l$ tokens. Each token is embedded and then passed through the encoder to obtain $D_l$-dimensional features. To facilitate subsequent computations, we repeat the language features $T$ times, thus resulting in the final language features $f_{t} \in \mathbb{R}^{T \times N_l \times D_l}$. %Done

For simplicity, we assume all the three encoders above use the same backbone, which means $D_v=D_l=D_a=D$. Without loss of generality, LSTTA allows using different backbones with varied dimensionality. %Done

\subsection{Basic Components}
Before presenting the two key modules in LSTTA, we first introduce two basic components: the cross-modal attention (CMA) unit and the adapter unit.  %Done
\vspace{3pt}
\\
\noindent\textbf{Cross-Modal Attention.}
%We adopt the self-attention in Transformer \cite{vaswani2017attention} to calculate the cross-modal attention for two groups of features from different modalities. To reduce the computational costs, 
We follow the strategy in \cite{LAVISH} to design a cross-modal attention (CMA) unit as follows\footnote{The CMA unit requires the dimensionality of $X$ and $Y$ to be equal. When this condition is not satisfied, we can simply add a linear projection layer before the CMA unit to make the dimensionality of the input features consistent. }: %Done
\begin{equation}
    \mathrm{CMA} (X,Y) = X + \mathrm{tanh} (g) \cdot \mathrm{Softmax} (XY^T)Y
\label{eq.2}
\end{equation}
where $g$ indicates a learnable scalar to balance the two terms. $\mathrm{tanh} (\cdot)$ is the tanh function to stabilize model training. 
\vspace{3pt}
\\
\textbf{Adapter.} Following previous work on adapter \cite{Adapter}, we use a two-layer MLP with bottleneck architecture, which consists of a down-projection layer $W_\mathrm{down}\in\mathbb{R}^{D\times D_h}$, a $\mathrm{ReLU}$ activation, and an up-projection layer $W_\mathrm{up}\in\mathbb{R}^{D_h \times D}$, where $D_h\ll D$ is a pre-defined hyper-parameter. Given an input feature $X$, the whole process of the adapter unit is defined as follows: %Done
\begin{equation}
\begin{split}
    \mathrm{Adapter}(X) = X + W_\mathrm{up} (\mathrm{ReLU} (W_\mathrm{down}(X))).
\end{split}
\end{equation}

\subsection{Long-Term Semantic Filtering Module}

The video semantics over time may not be uniform. Some video frames are meaningless or redundant. Therefore, it is necessary to filter out these uninformative frames. To this end, we introduce a long-term semantic filtering (LTSF) module, which is built upon the two basic components above, to calculate the semantic importance over time.

Taking the intermediate features $z_v$, $z_a$, $z_l$ from Eq. (\ref{eq.1}) as inputs, we first employ the CMA unit in Eq (\ref{eq.2}) to integrate information between two modalities. Specifically, we take the language features $z_l$ as the anchor and calculate cross-modal attention w.r.t. the visual features $z_v$ and audio features $z_a$, respectively:
\begin{equation}
\begin{split}
    z_{lv} = \mathrm{CMA} (z_l, z_v), \\
    z_{la} = \mathrm{CMA} (z_l, z_a).
\label{eq.3}
\end{split}
\end{equation}
where $z_{lv} \in \mathbb{R}^{T \times N_l \times D}$ and $ z_{la} \in \mathbb{R}^{T \times N_l \times D}$ are the attended features containing visual and audio semantics, respectively.

After obtaining the attended features $z_{lv}$ and $z_{la}$, we concatenate them along the token dimension to obtain the fused feature $z_\mathrm{lva} \in \mathbb{R}^{T \times 2N_t \times D}$ as follows:
\begin{equation}
    z_{lva} = \mathrm{Concat} (z_{lv}, z_{la}).
\label{eq.4}
\end{equation}

To measure the semantic importance over $T$ timestamps, we first perform mean-pooling on $z_{lva}$ along the token dimension and then feed the pooled feature through an adapter module to obtain the important scores $S\in\mathbb{R}^{T}$ as follows:
\begin{equation}
    S = \mathrm{Softmax} (\mathrm{Adapter} (\mathrm{MeanPool}(z_{lva}))).
\label{eq.5}
\end{equation}
The importance scores can be seen as a soft temporal mask over $T$ timestamps, which is further used in the short-term semantic interaction module to filter out irrelevant semantics. 

The flowchart of the LTSF module is shown in Fig.~\ref{fig:detail}(a).

\subsection{Short-Term Semantic Interaction Module}
Next, we introduce the short-term semantic interaction (STSI) module, which is also built upon the two basic components. As the visual and audio features are naturally aligned in the temporal dimension, it is essential to model their {token-level} short-term interaction within each timestamp. 

In contrast to the LTSF module which takes the language modality as the center, the STSI module takes the audio (or visual) modalities as the center and aggregates semantic information from the rest two modalities, resulting in the VL2A (or AL2V) variant. For simplicity, we use the VL2A variant as an example to explain the whole process of the STSI module. The flowchart of the STSI module is shown in Fig.~\ref{fig:detail}(b).   

Taking $z_v$, $z_a$, $z_l$ from Eq. (\ref{eq.1}) as inputs, a straightforward way to achieve the goal above is to concatenate $z_v$ and $z_l$ and then using the CMA unit to aggregate information from the concatenated features to $z_a$ as follows:
\begin{equation}
        \mathrm{CMA} (z_a, \mathrm{Concat} (z_{v},z_{l})).
    \label{eq.7}
\end{equation}
As the number of tokens in $z_a$ and $z_v$ is usually large, directly using CMA to integrate the audio and visual features may bring in non-negligible computational costs. Inspired by the success in previous work \cite{jaegle2021perceiver,LAVISH}, we use an alternative strategy that introduces a small set of $K$ learnable latent tokens $q \in \mathbb{R}^{T \times K\times D}$ to aggregate information from visual-language features into a small group of latent token features: 

\begin{equation}
        \widehat{q} = \mathrm{CMA} (q, \mathrm{Concat} (z_{a},z_{l}))
    \label{eq.8}
\end{equation}
where the output features $\widehat{q}\in\mathbb{R}^{T\times K\times D}$ contains condensed visual-language correlated semantics. Next, we aim to integrate $\widehat{q}$ with the audio feature $z_a$. Recall that we have obtained a temporal mask $S$ from the LTSF module to suppress uninformative timestamps. To make use of the temporal task, we slightly modify the CMA function and convert it to a masked CMA (MCMA) version as follows: 

\begin{table*}[t]
    \centering
    \scriptsize
    \setlength\tabcolsep{4.5pt}
    \caption{Performance comparison with state-of-the-art methods on Music-AVQA. $\dag$ means our re-implementation based on the official open-source code.}
    \begin{tabularx}{1.0\linewidth}{c|ccc|ccc|cccccc|c}
        \hline
        \multirow{2}*{Method} & \multicolumn{3}{c}{Audio Question (AQ)} \vline & \multicolumn{3}{c}{Visual Question (VQ)} \vline & \multicolumn{6}{c}{Audio-Visual Question (AVQ)} \vline & All\\ 
         ~ & Counting & Comparative & Avg. & Counting & Location & Avg. &Existential & Location & Counting & Comparative & Temporal & Avg. & Avg.  \\ \hline
        FCNLSTM~\cite{FCNLSTM} & 70.45 & 66.22 & 68.88 & 63.89 & 46.74 & 55.21 & 82.01 & 46.28 & 59.34 & 62.15 & 47.33 & 60.06 & 60.34 \\
        CONVLSTM~\cite{FCNLSTM} & 74.07 & 68.89 & 72.15 & 67.47 & 54.56 & 60.94 & 82.91 & 50.81 & 63.03 & 60.27 & 51.58 & 62.24 & 63.65 \\ 
        GRU~\cite{GRU} & 72.21 & 66.89& 70.24 & 67.72 & 70.11 & 68.93 & 81.71 & 59.44 & 62.64 & 61.88 & 60.07 & 65.18 & 67.07 \\
        HCAttn~\cite{HCAttn} & 70.25 & 54.91 & 64.57 & 64.05 & 66.37 & 65.22 & 79.10 & 49.51 & 59.97 & 55.25 & 56.43 & 60.19 & 62.30 \\
        MCAN ~\cite{MCAN} & 77.50 & 55.24 & 69.25 & 71.56 & 70.93 & 71.24 & 80.40 & 54.48 & 64.91 & 57.22 & 47.57 & 61.58 & 65.49 \\ 
        PSAC~\cite{PSAC} & 75.64 & 66.06 & 72.09 & 68.64 & 69.79 & 69.22 & 77.59 & 55.02 & 63.42 & 61.17 & 59.47 & 63.52 & 66.54 \\
        HME~\cite{HME} & 74.76 & 63.56 & 70.61 & 67.97 & 69.46 & 68.76 & 80.30 & 53.18 & 63.19 & 62.69 & 59.83 & 64.05 & 66.45 \\
        HCRN~\cite{HCRN} & 68.59 & 50.92 & 62.05 & 64.39 & 61.81 & 63.08 & 54.47 & 41.53 & 53.38 & 52.11 & 47.69 & 50.26 & 55.73 \\ 
        AVSD~\cite{AVSD} & 72.41 & 61.90 & 68.52 & 67.39 & 74.19 & 70.83 & 81.61 & 58.79 & 63.89& 61.52 & 61.41 & 65.49 & 67.44 \\
        Pano-AVQA~\cite{AVQA-yun2021pano} & 74.36 & 64.56 & 70.73 & 69.39 & 75.65 & 72.56 & 81.21 & 59.33 & 64.91 & 64.22 & 63.23 & 66.64 & 68.93 \\
        TG+SG~\cite{Music-AVQA} & 78.18 & 67.05 & 74.06 & 71.56 & 76.38 & 74.00 & 81.81 & 64.51 & 70.80 & 66.01 & 63.23 & 69.54 & 71.52 \\
        LAVISH~\cite{LAVISH} \dag & 75.59	&\textbf{84.13}	&76.86	&77.45	&72.91	&76.29	&71.91	&77.52	&75.81	&76.75	&77.62	&76.31	&76.10 \\ \hline
        \textbf{LSTTA (ours)} & \textbf{81.75}	&82.04	&\textbf{81.90}	&\textbf{81.82}	&\textbf{82.23}	&\textbf{82.03}	&\textbf{83.46}	&\textbf{79.11}	&\textbf{78.23}	&\textbf{78.02}	&\textbf{79.32}	&\textbf{79.63}	&\textbf{81.19} \\ \hline
    \end{tabularx}	
    \label{tab:music-avqa}

\end{table*}

\begin{equation}
    \mathrm{MCMA} (X, Y, S) = X + \mathrm{tanh} (g) \cdot \mathrm{Softmax} (S \otimes XY^T)Y.
\label{eq.9}
\end{equation}
where $\otimes$ refers to Hadamard product with broadcasting $S$ to the same shape of $X$. By using the MCMA unit, we obtained the attended audio features $\widehat{z}_{a}$ as follows:
\begin{equation}
    \widehat{z}_{a} = \mathrm{MCMA} (z_{a}, \widehat{q}, S)
   % z_{al} = \mathrm{MCMA} (z_{a}, \widehat{q}_{vl}, S).
\label{eq.10}
\end{equation}
where $\widehat{z}_{a}\in\mathbb{R}^{T\times N_a \times D}$ represents the attended audio features containing short-term interactions with the aligned visual features.

Similar to the LTSF module, we append an adapter unit on the attended audio features as follows:
\begin{equation}
    f_{a} = \mathrm{Adapter}(\widehat{z}_{a})
\label{eq.11}
\end{equation}
where $f_a$ corresponds to the output features in Eq. (\ref{eq.1}) of the current LSTTA block. By analogy, we can obtain the output visual features $f_v$ by simply exchanging the places of visual features $z_v$ and audio features $z_a$ in Eq. (\ref{eq.8})-(\ref{eq.11}). For the language modality, we directly feed the intermediate features $z_l$ as the output features $f_l$ of the LSTTA block in Eq. (\ref{eq.1}).

\section{Experimental Results}
We evaluate LSTTA on four trimodal learning datasets and perform a thorough comparison to the state-of-the-art methods of each dataset. Moreover, we conduct comprehensive ablation studies to explore the effectiveness of each component. %done

\subsection{Datasets and Evaluation Metrics}

\textbf{Music-AVQA}~\cite{Music-AVQA} consists of 9,288 videos featuring 22 different musical instruments, with a total duration of more than 150 hours. It contains 45,867 QA pairs covering three types of questions, namely AQ, VQ, and AVQ. AQ denotes audio-related questions, VQ denotes visual-related questions, and AVQ denotes audio-visual-related questions. For Music-AVQA, answer prediction accuracy is used as the standard metric for model evaluation.
\vspace{3pt}
\\
\noindent\textbf{CMU-MOSEI}~\cite{CMU-MOSEI} contains 23,456 videos and two subtasks: sentiment analysis and emotion recognition. The sentiment analysis task aims to predict the sentiment levels based on trimodal data. The emotion recognition task aims to predict one of the six emotion classes (happiness, sadness, anger, fear, disgust, or surprise). For the sentiment analysis task, several evaluation metrics are used jointly, including mean average error (MAE), Pearson correlation (Corr), binary classification accuracy (ACC-2), and F1 score. For the emotion recognition task, prediction accuracy (Acc) and F1 score are used as the evaluation metrics.
%With its large scale and rich annotations, the CMU-MOSEI dataset provides an excellent resource for advancing research on affective computing and multimodal sentiment analysis.
\vspace{3pt}
\\
\noindent\textbf{UR-FUNNY}~\cite{UR-FUNNY} collects 1,866 videos from the TED portal, covering 417 topics from 1,741 different speakers. Based on the laughter makeup, 8,257 humorous punchlines from the transcripts are chosen as positive examples, and 8,257 negative samples are chosen at random intervals. Prediction accuracy is used as the evaluation metric.
\vspace{3pt}
\\
\noindent\textbf{VIOLIN}~\cite{VIOLIN} collects 95,322 video-statement pairs from 15,887 video clips. Each video clip is paired with 6 statements and has an average length of 35.2s. All these videos are selected from popular TV shows and movies from YouTube channels. VIOLIN provides a multimodal inference task, which requires the model to predict whether a statement is true or false. %done
%\vspace{3pt}
%\\
%\noindent\textbf{Evaluation Metrics.} For the Music-AVQA tasks, we utilize answer prediction accuracy as the evaluation metric for our proposed model. . % Done

\subsection{Implementation Details}
We adopt the pre-trained CLIP model \cite{CLIP} as the default image encoder (ViT-L backbone with $N_v$=196, $D_v$=1,024) and language encoder (24-layer Transformer backbone with $N_l$=77, $D_v$=1,024), and w2v-Conformer~\cite{Conformer} as the default audio encoder (ViT-L backbone with $N_a$=196, $D_a$=1,024). Unless otherwise noted, we keep all the encoder parameters remain frozen, the number of latent tokens $K$=64, number of timestamps $T$=32, and latent dimensionality in the adapter unit $D_h$=512. 
All the models are trained using the AdamW optimizer~\cite{AdamW} on 8 Nvidia V100 GPUs. The cosine learning rate decay strategy with an initial learning rate of 8e-5. %done

\subsection{Main Results}
In this section, we conduct a series of experiments to compare with previous state-of-the-art methods.
\vspace{3pt}
\\
\noindent\textbf{Resutlts on Music-AVQA.} 
The results in Tab.~\ref{tab:music-avqa} show that LSTTA significantly outperforms previous state-of-the-art trimodal learning methods in terms of overall and per-type accuracies (AQ, VQ, and AVQ), verifying the superiority of our carefully-designed adapter architecture. Compared with the audio-visual bimodal adapter method LAVISH \cite{LAVISH}, LSTTA obtains a 5-point improvement, showing the effectiveness of trimodal modeling.
\begin{table}
    \centering
    \small
    \setlength\tabcolsep{6pt}
    \caption{Performance comparison with state-of-the-art methods on the sentiment analysis task of CMU-MOSEI.}
    % \vspace{-0.3cm}
    \begin{tabularx}{1.0\linewidth}{c|cccc}
    \hline
        Method & MAE ($\downarrow$) & Corr. & Acc-2 & F1-Score \\ \hline
        MuIT~\cite{MUIT} & 0.591 & 69.4 & -/81.6 & -/81.6 \\
        ICCN~\cite{ICCN} & 0.565 & 71.3 & -/84.2 & -/84.2 \\
        MISA~\cite{MISA} & 0.555 & 75.6 & 83.6/85.5 &83.8/85.3 \\
        ScaleAVLD~\cite{ScaleVALD} & 0.527 & 78.1 & 84.5/86.4 & 84.7/86.3 \\
        Self-MM~\cite{SelfMM} & 0.530 & 76.5 & 82.8/85.2 & 82.5/85.3 \\
        i-Code~\cite{i-code} & 0.502 & 81.1 & 85.3/87.5 & 85.6/87.4 \\ \hline
        \textbf{LSTTA (ours)} & \textbf{0.496} & \textbf{81.4} & \textbf{85.5}/\textbf{87.8} &\textbf{85.9}/\textbf{87.6}  \\     
    \hline
    \end{tabularx}	
    \label{tab:cmu-mosei-sentiment}
    % \vspace{-0.3cm}
\end{table}
\begin{table}
    \centering
    \small
    \setlength\tabcolsep{8pt}
    \caption{Performance comparison with state-of-the-art methods on the emotion recognition task of CMU-MOSEI. }
    % \vspace{-0.3cm}
    \begin{tabularx}{1.0\linewidth}{c|cccc}
    \hline
        Method & Acc. & F1-Score & Precision & Recall \\ \hline
        DFG~\cite{DFG} & 38.6 & 49.4 & 53.4 & 45.6 \\
        MISA~\cite{MISA} & 39.8 & 45.0 & 37.1 & 57.1 \\
        RAVEN~\cite{RAVEN} & 40.3 & 51.1 & 63.3 & 42.9 \\
        %MuIT~\cite{MUIT} & 42.3 & 52.3 & 63.6 & 44.5 \\
        HHMPN~\cite{HHMPN} & 43.4 & 52.8 & 59.1 & 47.6 \\
        %TAILOR~\cite{TAILOR} & 46.0 & 52.9 & \textbf{63.9} & 45.2 \\
        SIMM~\cite{SIMM} & 41.8 & 48.4 & 48.2 & 48.6 \\
        ML-GCN~\cite{ML-GCN} & 43.7 & 52.4 & 57.3 & 48.2 \\
        i-Code~\cite{i-code} & 50.2 & 56.2 & 50.7 & \textbf{63.0} \\ \hline
        \textbf{LSTTA (ours)} &\textbf{50.8} &\textbf{56.6} &\textbf{63.4} &51.1 \\
    \hline
    \end{tabularx}	
    % \vspace{-0.3cm}
    \label{tab:cmu-mosei-emotion}
\end{table}
\vspace{3pt}
\\
\noindent\textbf{Results on CMU-MOSEI.}
As shown in Tab.~\ref{tab:cmu-mosei-sentiment} and \ref{tab:cmu-mosei-emotion}, LSTTA achieves slightly better or at least comparable performance to the counterparts in both sentiment analysis and emotion recognition tasks of CMU-MOSEI. For the sentiment analysis task, LSTTA outperforms all the compared methods in terms of all criteria. Note that LSTTA achieves slightly better performance than the i-Code model pre-trained on the large-scale trimodal corpus ~\cite{i-code}, showing the effectiveness and efficiency of our method. 
For the emotion recognition task, LSTTA steadily outperforms all its counterparts in terms of most metrics. 
\begin{table}
    \centering
    \small
    % \scriptsize
    \setlength\tabcolsep{10pt}
    \caption{Performance comparison with state-of-the-art methods on UR-FUNNY and VIOLIN.}
    % \vspace{-0.3cm}
    \begin{tabularx}{0.8\linewidth}{c|cc}
        \hline
        Method & UR-FUNNY & VIOLIN \\ \hline
        MISA~\cite{MISA} & 70.61 & - \\
        MuIT~\cite{MUIT} & 70.55 & - \\
        HERO~\cite{HERO} & - & 68.59 \\
        GVE~\cite{GVE} & - & 68.39 \\
        i-Code~\cite{i-code} & \textbf{79.17} & 72.90 \\ \hline
        \textbf{LSTTA (ours)} & 79.10 & \textbf{73.06} \\
        \hline
    \end{tabularx}	
    \label{tab:ur}
    % \vspace{-0.3cm}
\end{table}
\vspace{3pt}
\\
\noindent\textbf{Results on UR-FUNNY and VIOLIN.}
To validate the generalization of our method, we conduct experiments on two more trimodal datasets UR-FUNNY and VIOLIN. The results in Tab.~\ref{tab:ur} suggest that LSTTA achieves superior or comparable performance to the compared methods. It is worth noting that all the compared methods are not adapter-based methods, which means they require much more training data and have more trainable model parameters. %done

% TODO: color bar 
\begin{figure*}
    \begin{center}
    	\includegraphics[width=0.97\linewidth] {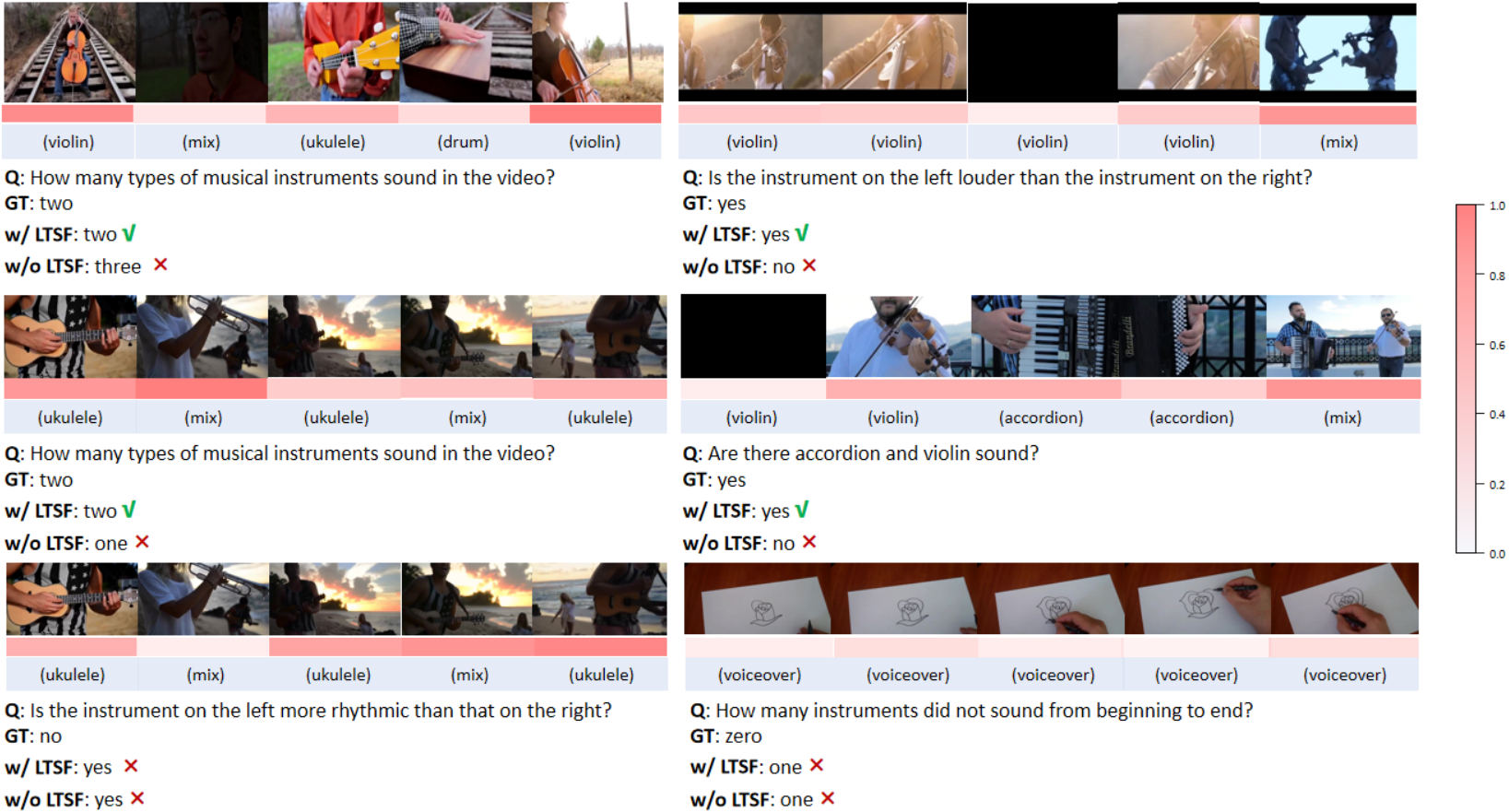}
    \end{center}
    % \vspace{-20pt}
    \caption{Visualization of six examples from Music-AVQA with questions (Q), ground-truth answers (GT), and prediction from LSTTA (w/ or w/o LTST module). For better understanding, we translate the audio contents into descriptions. The learned important scores by the LTSF module (see Eq. (\ref{eq.4})) are quantized into colored bins.} %done
    \label{soft-attn}
    % \vspace{-0.3cm}
\end{figure*}
% The visualization of soft attention mask. 

\subsection{Ablation Studies}
\begin{figure}
    \begin{center}
    \includegraphics[width=0.98\linewidth] {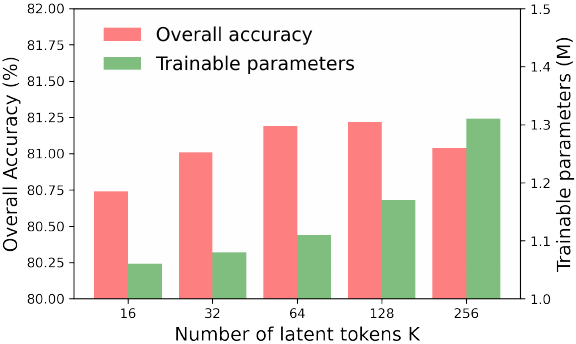}
    \end{center}
    \caption{Ablation of the number of the latent tokens $K$ in the STSI module. Ranging $K$ from 16 to 256, we report the overall accuracy and trainable parameters in one STSI block.}
    \label{fig:latent}
\end{figure}

We perform a series of ablation experiments on Music-AVQA to validate the effectiveness of the key components in LSTTA. The results are shown in Tab. \ref{tab:ablation}-\ref{tab:freeze} and Fig. \ref{fig:latent} and described next.
\vspace{3pt}
\\
\noindent\textbf{Effects of the STSI module. } 
As shown in Tab.~\ref{tab:ablation}, we first set a baseline model without introducing any LSTTA module (\#1). This results in a simple late-fusion model with independent and frozen encoders, which relies on task-specific heads to adapt to different trimodal learning tasks.
Based on the baseline model, we introduce two STSI modules AL2V and VL2A separately (\#2 and \#3) and observe distinct performance improvements on all question types. Besides, AL2V brings a larger gain on the AQ type while VL2A brings a larger gain on the VQ type (\#2 vs. \#3), which can be explained that the representations of the target modality have been enhanced by aggregated short-term semantic information from the rest modalities. Finally, the synergy of the two modules results in further improvement on all question types, especially for the AVQ type (\#4), reflecting the synergistic effect of the AL2V and VL2A modules in LSTTA.       
\vspace{3pt}
\\
\noindent\textbf{Effects of the LTSF module.} 
From the results shown in \#5 of Tab. \ref{tab:ablation}, we can see that introducing LTSF brings a significant improvement over the variant in \#4 in terms of all question types, especially for the question types containing the audio modality (2.7- and 2.6-points improvement are obtained on the AQ- and AVQ types, respectively). The result indicates that noise and irrelevant information is prevalent in the visual and audio modalities. The imposed long-term semantic filtering module effectively suppresses these meaningless timestamps and facilitates the representation capacity of the fused features in the STSI module. 
\begin{table}
    \centering
    \small
    % \scriptsize
    \setlength\tabcolsep{6pt}
    \caption{Ablations of key modules of LSTTA on Music-AQVA. LTSF refers to the long-term semantic filtering module. AL2V and VL2A refer to two short-term semantic interaction modules. LT means the latent tokens.}
    %\vspace{-0.3cm}
    \begin{tabularx}{1.0\linewidth}{c|cccc|ccc}
        \hline
         & LTSF & AL2V & VL2A & LT & AQ & VQ & AVQ \\ \hline
        1 & \XSolidBrush & \XSolidBrush & \XSolidBrush & \XSolidBrush &76.71 & 79.56 &65.73 \\
        2 & \XSolidBrush & \Checkmark & \XSolidBrush & \XSolidBrush &76.79 & 80.07 &67.57 \\
        3 & \XSolidBrush & \XSolidBrush & \Checkmark & \XSolidBrush &77.54 & 79.67 &66.17 \\
        4 & \XSolidBrush & \Checkmark & \Checkmark & \XSolidBrush &78.64 & 80.23 &75.91 \\
        5 & \Checkmark & \Checkmark & \Checkmark & \XSolidBrush &81.34 & 81.89 &78.53 \\
        6 & \Checkmark & \Checkmark & \Checkmark & \Checkmark &\textbf{81.90} & \textbf{82.03} &\textbf{79.63} \\
        \hline
    \end{tabularx}	
    \label{tab:ablation}
    % \vspace{-0.3cm}
\end{table}

To better verify the effectiveness of the LTSF module, we visualize the learned temporal mask for several typical examples in Fig.~\ref{soft-attn}. From the results, we can see that the learned temporal mask can filter out meaningless or uninformative timestamps in most cases, thus improving the answering accuracy. However, there are still some failure cases (the last row) on subjective questions or complex background music. These observations inspire us to introduce more powerful language and audio encoders in the future.
\vspace{3pt}
\\
\noindent\textbf{Effects of the latent tokens in STSI.} 
As shown in \#6 of Tab. \ref{tab:ablation}, we can see that introducing latent tokens not only reduces the computational costs but also brings performance improvement. Next, we investigate the effect of the number of latent tokens to model performance. The results in Fig.~\ref{fig:latent} show that as the increase of the number of latent tokens $K$, the model performance is nearly saturated at $K$=64. Further increasing $K$ brings marginal improvement or even degradation at the expense of much more parameters and computational costs. To make a trade-off, we set the number of latent tokens to 64 in our experiments. 
\vspace{3pt}
\\
\noindent\textbf{Effects of the different modality encoders.} 
As mentioned above that LSTTA supports different modality encoders, we explore the effects of different modality encoders in Tab.\ref{tab:bakcbone}. From the results, we can see that replacing the visual encoder (\#2 vs. \#1) or language encoder (\#3 vs \#1) does not bring performance drops. This suggests that multimodal alignment can be effectively established by using LSTTA, regardless of whether the modality encoders have been pre-trained jointly or not (i.e., CLIP-V and CLIP-T). Besides, as the capability of the language encoder in CLIP is not strong enough, replacing it with a more powerful counterpart, e.g., BART \cite{BART}, brings performance improvement. Similar observations have been witnessed in previous work ~\cite{METER}.

\begin{table}
    \small
    \centering
    \setlength\tabcolsep{3.5pt}
    \caption{Ablations of using different backbones for different modalities, including the CLIP-V \cite{CLIP} and Swin \cite{Swin-transformer} visual backbones, CLIP-T \cite{CLIP} and BART \cite{BART} language backbones, and W2V-Conformer \cite{Conformer} backbone.}
    \begin{tabularx}{1.0\linewidth}{c|ccc|cccc}
        \hline
         & Visual & Lang. & Audio & AQ & VQ & AVQ & All\\ \hline
        1 & CLIP-V & CLIP-T & W2V-Conf & 81.90 & 82.03 & 79.63 & 81.19\\ 
        2 & Swin & CLIP-T & W2V-Conf & 81.67 & 81.74 & 79.28 & 80.90\\
        3 & CLIP-V & BART & W2V-Conf & \textbf{82.02} & \textbf{82.05} & \textbf{79.76} & \textbf{81.28}\\
        \hline
    \end{tabularx}	
    \label{tab:bakcbone}
\end{table}
\vspace{3pt}

\noindent\textbf{Effects of parameter-efficient learning.} 
In Tab.~\ref{tab:freeze}, we compare the effectiveness and parameter-efficiency of LSTTA (with frozen modality encoders) and the counterpart with full-parameter fine-tuning. From the results, we can see that the LSTTA has $1/14\times$ fewer trainable parameters while achieving a 2.5-point higher accuracy than the fully fine-tuned variant. This can be explained that the goal of trimodal learning lies in establishing interaction across modalities, which is different from the objectives of the pre-trained modality encoders and may meet the catastrophic forgetting problem~\cite{kirkpatrick2017overcoming,Adapter}. 

\begin{table}
    \centering
    \small
    \setlength\tabcolsep{6.5pt}
    \caption{The ablation of parameter-efficiency learning.}
    \begin{tabularx}{1.0\linewidth}{c|ccccc}
        \hline
         & \makecell{Trainable\\params (M)} & AQ & VQ & AVQ & All\\ \hline
        freeze enc. & 83M & \textbf{81.90}& \textbf{82.03} & \textbf{79.63} & \textbf{81.19}\\
        fully f.t. & 1,150M & 79.34 & 81.27 &75.28 & 78.63 \\
        \hline
    \end{tabularx}	
    \label{tab:freeze}
\end{table}

\section{Conclusion}
In this paper, we propose a novel Long Short-Term Trimodal Adapter (LSTTA) method for universal trimodal learning tasks. Unlike prior works requiring large-scale pre-training or fine-tuning the entire model, LSTTA aims to utilize pre-trained unimodal or bimodal encoders and introduce an adapter architecture with very few parameters to adapt to downstream tasks effectively and efficiently. LSTTA consists of a long-term semantic filtering module and two short-term semantic interaction modules. Results on four trimodal learning datasets show that LSTTA achieves superior performance compared with state-of-the-art methods.  
We hope our study may serve as a baseline to inspire future research on trimodal learning and beyond.

\noindent \textbf{Acknowledgement}
This work was supported in part by the National Natural Science Foundation of China under Grant No.62072147 and in part by the Zhejiang Provincial Natural Science Foundation of China under Grant No.LR22F020001 and No.LDT23F02025F02.

\clearpage
{\small
\bibliographystyle{ieee_fullname}
\bibliography{egbib}
}

\end{document}